\documentclass{sigchi}

\usepackage{balance}  
\usepackage{graphics} 
\usepackage[T1]{fontenc}
\usepackage{txfonts}
\usepackage{mathptmx}
\usepackage[pdftex]{hyperref}
\usepackage{color}
\usepackage{booktabs}
\usepackage{textcomp}
\usepackage{microtype} 
\usepackage{ccicons}  
\usepackage[utf8]{inputenc} 

\usepackage{todonotes}

\def\plaintitle{Calendar.help: Designing a Workflow-Based \\ Scheduling Agent with Humans in the Loop}

\def\plainauthor{Justin Cranshaw, Emad Elwany, Todd Newman, Rafal Kocielnik, Bowen Yu, Sandeep Soni, Jaime Teevan, Andrés Monroy-Hernández}
\def\plainkeywords{Scheduling; microtask; macrotask; crowdsourcing; conversational agent; assistant.}

\makeatletter
\def\url@leostyle{%
  \@ifundefined{selectfont}{
    \def\UrlFont{\sf}
  }{
    \def\UrlFont{\small\bf\ttfamily}
  }}
\makeatother
\def\pprw{8.5in}
\def\pprh{11in}

\definecolor{linkColor}{RGB}{6,125,233}
\hypersetup{%
  pdftitle={\plaintitle},
  pdfauthor={\plainauthor},
  pdfkeywords={\plainkeywords},
  bookmarksnumbered,
  pdfstartview={FitH},
  colorlinks,
  citecolor=black,
  filecolor=black,
  linkcolor=black,
  urlcolor=linkColor,
  breaklinks=true
}

\toappear{\scriptsize Permission to make digital or hard copies of part or all of this work for personal or classroom use is granted without fee provided that copies are not made or distributed for profit or commercial advantage and that copies bear this notice and the full citation on the first page. Copyrights for third-party components of this work must be honored. For all other uses, contact the owner/author(s). Copyright is held by the author/owner(s). \\
{\emph{CHI 2017, May 6-11, 2017, Denver, CO, USA.} } \\
ACM ISBN 978-1-4503-4655-9/17/05. \\
http://dx.doi.org/10.1145/3025453.3025780}

\begin{document}

\title{\plaintitle}

\numberofauthors{1}
\author{
  \alignauthor{Justin Cranshaw$^1$, Emad Elwany$^1$, Todd Newman$^1$, Rafal Kocielnik$^2$,\\Bowen Yu$^3$, Sandeep Soni$^4$, Jaime Teevan$^1$, Andrés Monroy-Hernández$^1$\\
   \affaddr{$^1$ Microsoft Research, Redmond, WA, USA}\\
    \affaddr{$^2$ University of Washington, Seattle, WA, USA}\\
    \affaddr{$^3$ University of Minnesota, Minneapolis, MN, USA}\\
    \affaddr{$^4$ Georgia Institute of Technology, Atlanta, GA, USA}\\
    }
}

\maketitle

\begin{abstract}
%
Although information workers may complain about meetings, they are an essential part of their work life. Consequently, busy people spend a significant amount of time scheduling meetings. 
We present Calendar.help, a system that provides fast, efficient scheduling through structured workflows. Users interact with the system via email, delegating their scheduling needs to the system as if it were a human personal assistant. Common scheduling scenarios are broken down using well-defined workflows and completed as a series of microtasks that are automated when possible and executed by a human otherwise. Unusual scenarios fall back to a trained human assistant who executes them as unstructured macrotasks. We describe the iterative approach we used to develop Calendar.help, and share the lessons learned from scheduling thousands of meetings during a year of real-world deployments. Our findings provide insight into how complex information tasks can be broken down into repeatable components that can be executed efficiently to improve productivity.
\end{abstract}

\category{H.5.m.}{Info. interfaces and presentation (e.g., HCI)}{Misc.}


\section{INTRODUCTION}

Scheduling a meeting can feel like a frustrating distraction from the things that matter, so much so that some professionals hire assistants to help with the task. Many others would like to have scheduling help, but cannot afford a full time assistant and thus turn to software solutions. Most calendaring tools, however, are too rigid to handle the wide variation in people's meeting needs, and are particularly cumbersome when attendees need to coordinate multiple schedules while each using their own individual tool and approach.

We present Calendar.help, a service that allows people to schedule meetings as if they were working with a human assistant while automating and structuring many aspects of the task behind the scenes. Calendar.help subscribers interact with the system by emailing Cal, their virtual assistant. For example, if Bob, a Calendar.help subscriber, wanted to meet with Alice, he could email her and cc Cal: 
\begin{quote}
``Alice, can we get together sometime next week to discuss the CHI reviews? Cc'ing Cal to help us find a time. ---Bob''
\end{quote} 
Cal, the assistant, would then follow up directly with Alice, handling all of the back-and-forth emails necessary to find an acceptable meeting time and finalize the details (e.g., location, contact information, etc.). Cal would then create a meeting on the subscriber's calendar and automatically send an invitation to Alice. 
By building on the familiar practice of copying an assistant when asking for help,  
Calendar.help provides a useful service that does not force people to change how they currently work or interact with others.

\begin{figure}
\centering
  \includegraphics[width=\columnwidth]{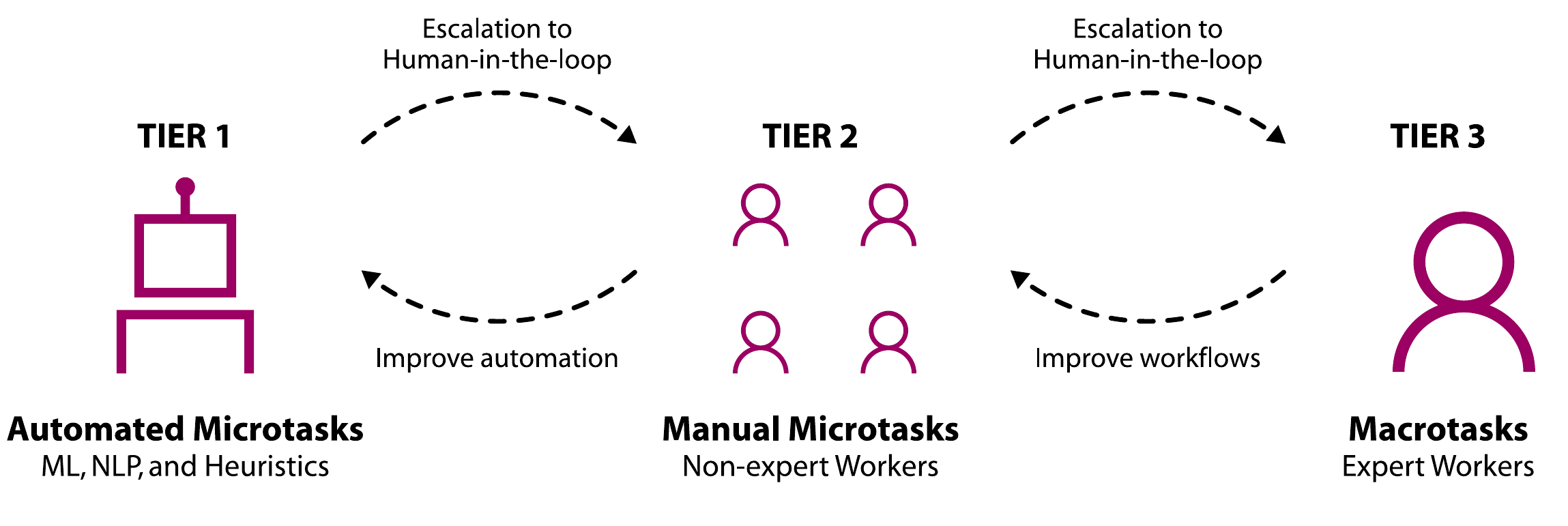}
  \caption{A three-tiered architecture for efficiently handling tasks. When possible, macrotasks (Tier 3) are decomposed into a workflow of smaller dependent microtasks that are executed manually by non-expert workers (Tier 2) and automated when possible (Tier 1).}~\label{fig:architecture}
\end{figure}

Calendar.help efficiently and robustly handles a broad range of calendaring needs by structuring and automating incoming requests whenever possible. As shown in Figure~\ref{fig:architecture}, task execution is organized into three tiers. Common scheduling tasks
are broken down using well-defined workflows into a series of microtasks (e.g., identifying the location of a meeting) that are executed automatically (Tier 1) or manually (Tier 2). Uncommon or complex scenarios, such as re-scheduling, are done by a scheduling expert as macrotasks (Tier 3). 
This approach makes the execution of simple requests efficient, and the execution of complex requests possible.

The main contributions of this work are the design, deployment, and study of Calendar.help, a production-quality system used by hundreds of people to schedule thousands of real-world meetings for a year. Unlike earlier approaches, the system supports people's existing calendaring tools and idiosyncrasies by adopting email as its primary user interface and conversation as its primary mode of interaction. We introduce a novel architecture that seamlessly combines automation, microtasking, and macrotask execution to produce a responsive and scalable scheduling assistant, and demonstrate its value through large-scale field deployments.

\section{RELATED WORK}

Scheduling meetings is a difficult task that requires communication, coordination, and negotiation among multiple parties \cite{grudin1988cscw}, where each person may have conflicting availabilities, and each may use different, non-interoperable tools \cite{bardram2005web}. The negotiation is itself complex, as meeting times and locations are resources often governed by intricate constraints and external dependencies \cite{sen1998formal}. Furthermore, the parties may be geographically dispersed, introducing additional constraints due to time zone differences and the need for communication technologies for remote meetings \cite{tang2011your}. This negotiation often needs to occur asynchronously, sometimes requiring several days for the parties to reach consensus \cite{Ehrlich1987social, ehrlich1987strategies}. Once a meeting is scheduled, it needs continuous maintenance, as new events often prompt meeting updates and re-schedules.  

Some people hire assistants to delegate their scheduling needs \cite{Ehrlich1987social}, and, in a study of administrative assistants, all but one reported that most of their work was scheduling related \cite{Erickson2008assistance}. 
But although research shows that people are more productive when they delegate tasks such as scheduling and replace them with higher-value ones \cite{birkinshaw2013make}, most people cannot afford to hire an assistant. The vast majority of information workers resort to scheduling their meetings themselves using a variety of tools. Some people use shared calendars to mitigate the burden of coordinating meetings \cite{Reinecke2013doodle}, but this requires that attendees share their availability openly \cite{Kellermann2009privacy} or use compatible calendaring systems \cite{Myllarniemi2014meeting}. Further, popular web-based coordination tools like Doodle face challenges around participants' social norms and values \cite{Reinecke2013doodle}. 
Despite the myriad of scheduling tools available, a survey of information workers found that 80\% of people use email for scheduling \cite{ducheneaut2001mail}. This and other findings have led researchers to stress the need to integrate the communication and calendaring functions of digital tools \cite{Ehrlich1987social}.

Artificial intelligence researchers have explored automated approaches to identify mutually convenient times for meetings \cite{brzozowski2006grouptime}, and, more broadly, to delegate tasks commonly done by assistants to software agents \cite{horvitz1999principles, maes1994agents, mitchell1994experience, myers2007intelligent, zunino2009chronos}. However, it has been difficult for these systems to capture the nuances of what people want when they schedule meetings, limiting the usefulness of these completely-automated approaches. 
A hybrid intelligence approach is an alternative that leverages human and machine intelligence, often combining microtask workflows with automation \cite{hybridintelligence}. 

Microtask workflows are being increasingly used to accomplish complex, multi-step tasks \cite{kittur2013future}, such as taxonomy-creation \cite{chilton2013cascade}, itinerary-planning \cite{zhang2012human}, writing \cite{teevan2016supporting}, and even real-time conversations \cite{lasecki2013chorus}. Workflows have even been used to assemble flash teams of expert workers of different specialties  \cite{retelny2014Expert}.
Prior work has demonstrated several advantages to breaking a monolithic task into microtasks, including making tasks easier for workers to complete~\cite{teevan2016microwork}, producing higher quality outcomes~\cite{Cheng2015break}, reducing coordination and collaboration overheads~\cite{teevan2016supporting}, and facilitating partial automation of a larger task \cite{kamar2012combining}. 
Emerging services such as Facebook M \cite{wired_facebookM}, X.ai and Clara Labs \cite{bloomberg_chatbots, wired_ai_helps_humans} use a combination of automation and human labor while providing users with a seamless experience. However, very little has been shared about how these systems actually work and are used. 




Bardram and Bossen argue that it is difficult for a single information system to support the `web of coordinative artifacts' that people employ to schedule their meetings \cite{bardram2005web}. Our goal with Calendar.help is not to obviate this network of interdependent scheduling tools and processes, but rather to introduce a flexible and adaptive virtual assistant that can navigate it on a person's behalf using a conversational approach over email. We accomplish this with a novel architecture that seamlessly combines automation, structured microtask workflows, and unstructured macrotasks.

\section{THE CALENDAR.HELP SYSTEM}


\subsection{User Experience}

People subscribe to Calendar.help at \url{http://calendar.help} by providing their email address and some default meeting preferences, such as their preferred meeting length. They also authenticate their Microsoft or Google accounts using oAuth\footnote{A standard for delegated access to resources: \url{https://oauth.net/}.} to grant the system permission to access their online calendars. Subscribers then primarily interact with the system through email by including the Calendar.help virtual scheduling assistant in meeting-related email conversations. While we used a variety of names for the assistant throughout our user studies, for simplicity we refer to it as \emph{Cal} in this paper.

The Calendar.help system engages when a subscriber emails or cc's Cal at {\tt cal@calendar.help}.
When Cal first learns about a meeting, it identifies the meeting constraints based on information provided in the originating email message, and the subscriber's calendar availability and meeting preferences.
Given these constraints, Cal then sends a `ballot email' to invitees with potential meeting times  available in the subscriber's calendar (e.g., ``\emph{Alice, I checked Bob's calendar and these times are available: Wed Sep 20 at 9am, [...] ---Cal}'').

If none of the proposed time options work, Cal continues to negotiate with the attendees to determine other time options until the meeting is scheduled.
Cal follows up with invitees by sending up to two reminders at appropriate intervals if invitees have not replied to the ballot.

Once invitees reply to Cal with their choice (e.g., ``\emph{Great. Wednesday at 9 works! ---Alice}''), Cal adds the meeting to the subscriber's calendar, and sends a notification email to the subscriber and the invitees. The message to the invitees contains a machine-readable invitation\footnote{An .ics file: \url{https://tools.ietf.org/html/rfc5545}.}  they can use to add the event to their preferred calendaring tool.




Cal aims to reduce the burden of reading unnecessary email by communicating directly with individual people rather than emailing everyone at once.  Although Cal may send many emails in the process of scheduling a meeting, individuals only see a small slice of the total communication. Even complex scenarios follow this principle. When coordinating with multiple attendees, handling failed ballots, rescheduling or cancelling a meeting, or collecting contact information from invitees, people are only included when required.

\begin{figure}
\centering
  \includegraphics[width=\columnwidth]{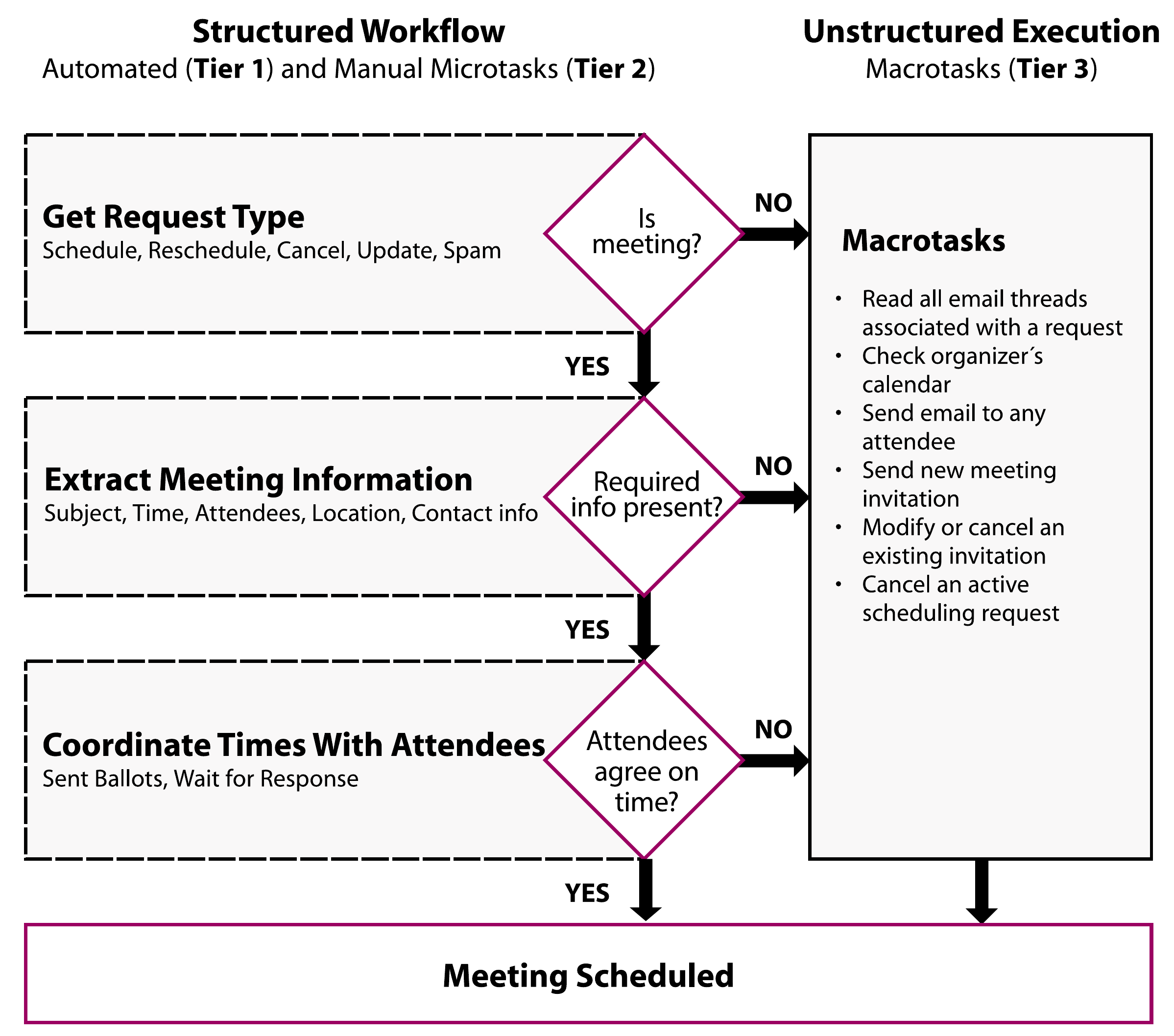}
  \caption{A simplified version of the  structured workflow's main phases: determining the type of request, extracting essential information, coordinating times and scheduling the meeting. At each point, lack of essential information or unexpected behavior can be handled in macrotasks.}~\label{fig:workflow}
\end{figure}

\subsection{Architecture}

A central component of Calendar.help is a scheduling workflow that specifies all the interactions between Cal, the meeting organizer, and the invitees, encapsulating our domain knowledge acquired during the iterative design of the system.
One of the main goals of the workflow is to gather information about the meeting from the attendees to ensure it is scheduled as requested and according to everyone's availability. In doings so, the workflow encodes several aspects for processing a scheduling request, including the timing and content of Cal's emails to attendees, and procedures for extracting information from attendee responses to determine when to schedule a meeting.

In order for Cal to participate in complex communications about scheduling with high reliability, we designed a three-tiered architecture that balances access to skilled human assistance as needed, with a path towards automation through \emph{microtasking}, as illustrated in Figure~\ref{fig:architecture}. The workflow specifies rules for how requests are handled across these three tiers, and processes them in two modes: as \emph{structured microtasks} (Tiers 1 and 2), and \emph{unstructured macrotasks} (Tier 3).  In Tiers 1 and 2, each atomic information gathering component is implemented as a \emph{microtask}, a small, self-contained unit of work with a fixed input (e.g., an email message, or the organizer's calendar availability), and a fixed output (e.g., the meeting duration, or a phone number). In Tier 3, the scheduling \emph{macrotask} is executed by an expert worker, empowered to schedule the meeting following certain guidelines and their own expertise.


Each microtask is automated when possible by the machine learning, natural language processing, or heuristic-based components
contained in Tier 1. 
If these automated components produce low-confidence results, or if there is no automation in place at all for a particular microtask, the workflow allocates the microtask to a non-expert human worker in Tier 2.
If this non-expert worker cannot complete the microtask because, for example, the task input does not match what is being asked, or the scenario at hand is still unaccounted for in the workflow, the request is packaged as a macrotask and sent to an expert worker in Tier 3.

This architecture allows task execution to be escalated to higher, more capable, and generally more expensive tiers as the lower tiers fail.
Figure \ref{fig:workflow} shows a simplified example of a workflow used by Calendar.help with several escalation points highlighted.
The main objective of this design is to maintain a reliable user experience from day one, while affording efficiency gains over time, as historical usage data can directly inform future designs to mitigate escalations.
New microtasking workflows can be designed by analyzing historical macrotask escalation points and reducing the frequency of macrotask escalations in future versions.
Additionally, as workers execute microtasks, the system collects a corpus of labeled data to train new machine learning components and improve existing heuristics for automating microtasks.

\subsubsection{Tier 1: Automatic Microtask Execution}

The automatic execution of microtasks relies on a combination of machine learning (ML), natural language processing (NLP), and heuristic-based components. 
Well-studied NLP techniques such as intent detection, entity extraction \cite{etzioni2005unsupervised, angeli2012parsing}, and slot filling \cite{xu2013convolutional} are suitable for processing the email associated with a microtask and gathering relevant information from emails to automate microtasks, such as the meeting duration or time options.
ML can also be used to model a microtask, predicting the task output given its input and training data, and also for modelling confidence estimates of various inferences the system makes, driving workflow decision making across tiers.
Additionally, heuristics can be used to automate commonly occurring microtask scenarios, for example, when Cal receives an email, the system attempts to automatically determine whether the message relates to an existing request or a new one using a set of hard-coded heuristics, including, searching properties of the email header.

The initial versions of Calendar.help discussed in this paper have limited ML or NLP-based automation deployed in production settings, which require high precision to meet users' demands for quality. In this work we highlight initial results that illustrate how data collected from the system can bootstrap initial ML and NLP-based automation efforts.


\subsubsection{Tier 2: Manual Microtask Execution}


The microtasks that Calendar.help cannot execute automatically are listed on an in-house online microtasking platform similar to Amazon Mechanical Turk.
For example, if Calendar.help fails to classify whether an incoming message relates to an existing meeting, a worker is asked to provide this classification.
All manual microtasks are presented with a consistent three-part design pattern. At the top is a section with instructions. The originating email message is shown on the left side, and actions (such as options to choose from or text fields to complete) are shown on the right. In the case of email classification, the worker will be given options that include, \emph{``The message is about a new meeting,''} or \emph{``The message is about an existing meeting.''}
The interface is designed to show only the information necessary to execute the specific microtask.
If a microtask cannot be completed using the information provided, workers are also given the option to click a button labeled, \emph{``I can't answer.''} This causes the task to escalate to Tier 3.
After completing a microtask, workers are given the next microtask in the queue.



\subsubsection{Tier 3: Manual Macrotask Execution}


Macrotasks are also listed as tasks on an in-house platform and assigned to workers with scheduling expertise.
The macrotask displays all of the messages related to the meeting that have been relayed between Cal and the invitees, any the information collected thus far, the organizer's anonymized calendar, and the details of the meeting invitation if one has already been sent. Workers are asked to use their skills to determine the best next course of action for the meeting and execute it.
Possible actions include sending a message to a participant asking for more details, sending the meeting invitation if all the necessary information has been collected, canceling the meeting, updating a meeting invitation with new details, or simply pushing the task back in the queue if a response is still required from a participant.
After completing a macrotask, workers are given the next macrotask in the queue.


For the sake of simplicity, in the studies discussed in this paper, we used the same worker pool for tasks in Tiers 2 and 3, though we maintained a separate task queue for each tier to maintain componentization of the work.
Additionally, instead of the market-place approach to online task completion where a task is listed and workers are paid per task completed, we hired workers in shifts, who continually monitor our task queues and are paid hourly.  
Workers are required to sign a non-disclosure agreement to ensure confidentiality. Future work in this area will address anonymization of user data, so that workers can complete scheduling tasks while maintaining user privacy.


\subsubsection{Implementation}

The management of tasks across tiers is executed by an event-driven workflow engine run in a distributed manner on Azure compute nodes, utilizing the Azure Service Bus distributed queue capabilities for firing and distributing events.
The engine manages the execution of workflow tasks by building a dependency graph and ensuring that these tasks are executed in the right order. It also handles running both synchronous and asynchronous tasks in a uniform manner, masking this complexity from the workflow developer.
The engine takes care of pausing workflow execution and persisting the state in Azure persistent stores
to free resources. Then, when the asynchronous output is available, which is indicated by an external event, the engine loads the state and resumes execution from the point at which it previously stopped.
The assistant's mailbox is managed using the Exchange Web Services API via Microsoft Bot Framework\footnote{\url{http://botframework.com}}.
As a means of facilitating iterative design improvements based on user feedback, workflow versioning allows the system to continue serving long running requests, like meeting requests that typically take days to complete, while maintaining the ability to deploy new versions with new functionality and otherwise breaking changes.
Older requests continue to be served by the older versions, while newer versions handle new requests.


\section{EVALUATION}

\begin{table}
  \centering\small
    \begin{tabular}{rp{1.3cm}p{1.3cm}p{1.3cm}p{1.3cm}}\toprule
      &\bfseries Study 1&\bfseries Study 2&\bfseries Study 3a&\bfseries Study 3b\\
      &\emph{Wiz. of Oz}&\emph{Usability}&\emph{Field}&\emph{Field}\\\midrule
      \bfseries Start date&4/6/2015&6/24/2015&11/4/2015&4/5/2016 \\
      \bfseries End date&4/26/2015&9/1/2015&3/27/2016&8/25/2016 \\\midrule
      \bfseries Duration&14 days&7 to 14 days&Open usage &Open usage \\
      \bfseries Gratuity & Yes & Yes & No & No \\
      \bfseries Workers&6&6&7&7 \\
      \bfseries Participants&11&24&65&178 \\\midrule
      \bfseries Invitees&100&61&1,030&1,981 \\
      \bfseries Meetings&146&124&918&1,626 \\
      \bfseries Emails&2,638&1,723&8,884&15,659 \\\bottomrule
  \end{tabular}
  \caption{Summary of our evaluations  of Calendar.help.}~\label{tab:studies}
\end{table}

We evaluated Calendar.help through a series of user studies conducted over the course of an 18-month period, summarized in Table~\ref{tab:studies}.
Each study assessed different aspects of the system at different levels of maturity. As we learned from our users' experiences we made iterative improvements.

\subsection{Study 1: Wizard of Oz Study}

\emph{Purpose:} Prior to implementing software, we conducted a Wizard of Oz study \cite{kelley1984iterative} to determine how people delegate scheduling to an email-based assistant in a naturalistic context. The goal was to generate initial specifications for the scheduling workflows and validate basic interaction assumptions.

\emph{Participants:} We recruited 11 ``busy information workers'' in various occupations, who had above-average scheduling demands, and who did not have access to an administrative assistant. Participants included a wedding florist and a small business owner in Seattle, a hospital administrator in California, and an emergency management program manager in Nebraska.
Over the course of two weeks, participants were required to schedule at least five meetings a week using Calendar.help, for which they received a \$100 USD gratuity.

\emph{System:} Each participant was given a unique email address for their assistant to include on scheduling requests (e.g. Alice's assistant was at {\tt meet.alice@calendar.help}). Emails sent to any of the 11 assistant addresses would forward to the same inbox, where a set of email rules sorted them into separate folders for each participant. We then hired six expert scheduling assistants through the freelancing platform Upwork\footnote{Available at \url{http://upwork.com}} who were given access to the central email inbox mentioned above. These experts were paid above U.S. minimum wage rates (\$15 USD or more per hour), depending on the market dynamics. Each participant's mailbox folder provided the full history of interactions for each participant and their meeting invitees. This helped workers isolate the context of the participant's meetings. Workers were staffed in shifts, sharing critical information across shifts through a system of folders, tags, and a shared online notebook.

\begin{table}
  \centering\small
    \begin{tabular}{lrrr}\toprule
      User & Num weeks & Meetings & Meetings per week \\ \midrule
1   &  13.4 & 180 & 13.4 \\
2   &  14.1 & 141 & 10.0 \\
3   &  18.3 & 128 & 7.0 \\
4   &  20.3 & 107 & 5.3 \\ 
5   &  18.1 & 106 & 5.8 \\
6   &  19.1 & 98 & 5.1 \\ 
7   &  17.9 & 71 & 3.0 \\
8   &  17.3 & 70 & 4.0 \\ 
9   &  16.0 & 61 & 3.8 \\ 
10  &  20.1 & 54 & 2.7 
      \\\bottomrule
  \end{tabular}
  \caption{Usage summary of the 10 most active users in Study 3b. Out of all 178 users, the average user scheduled 2.1 meetings per week over the course of the study.}~\label{tab:TopUsers}
\end{table}

\subsection{Study 2: Usability Study}

\emph{Purpose:} To test the viability of Calendar.help and to make basic design improvements, we conducted a usability study where we deployed a primitive version of the system built using the three-tiered architecture described already. 

\emph{Participants:} We recruited 24 participants, including some from the first study. Again, we tried to recruit information workers who had meeting scheduling needs. Participants received a \$100 USD gratuity, and were, once again, required to schedule at least five meetings a week using Calendar.help over the course of a one- or two-week study period. 


\emph{System:} Instead of using a shared inbox as in Study 1, the same six workers, again working in shifts, monitored a work queue and performed microtasks and macrotasks as the system generated them. In this study, all participants interacted with the same email address for the assistant (e.g. {\tt cal@calendar.help}), and the system took care of providing the necessary context to the workers as they completed each microtask and macrotask. 




\subsection{Study 3: Field Deployment}

\emph{Purpose:} We conducted a field deployment of Calendar.help to assess user engagement and system performance without the biases introduced by the gratuity and usage requirements of Studies 1 and 2. 

\emph{Participants:} 
We released Calendar.help within a large multinational technology corporation of approximately one hundred thousand employees.
Calendar.help was advertised internally through word of mouth, mailing lists, and online boards.
We actively recruited people in non-technical roles, such as those in recruiting, business development, sales, and marketing, since their meetings often included invitees outside the organization. 
Participants had to apply to join, which allowed us to select for these roles.



\emph{System:} During the field deployment there were two major Calendar.help releases. We refer to the period after the first release as Study 3a and the period after the second release as Study 3b. The user experience was largely consistent across releases. The system, however, was significantly more stable during 3b, as several bugs had been addressed, and it included a robust instrumentation infrastructure.
Study 3b included all participants from Study 3a plus new users who organically chose to adopt Calendar.help. In this study, we staffed the microtask and macrotask worker pools using a professional staffing company that provided wages significantly above minimum wage and benefits.




\begin{figure}
\centering
  \includegraphics[width=\columnwidth]{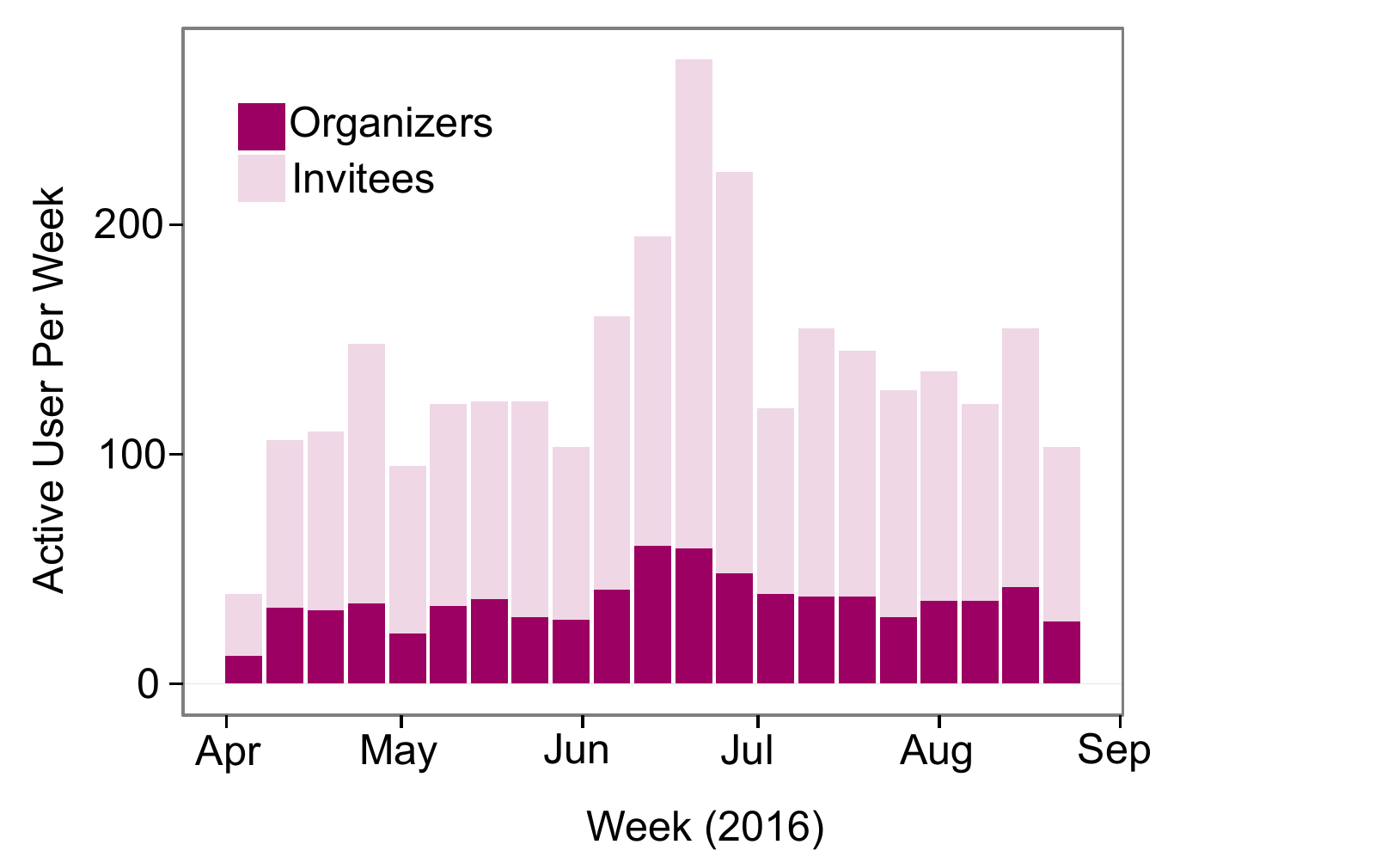}
  \caption{The total number of active users and invitees during each week of Study 3b.}~\label{fig:UsersPerWeek}
\end{figure}

\section{RESULTS}
\subsection{Usage Statistics}

\begin{figure}
\centering
  \includegraphics[width=\columnwidth]{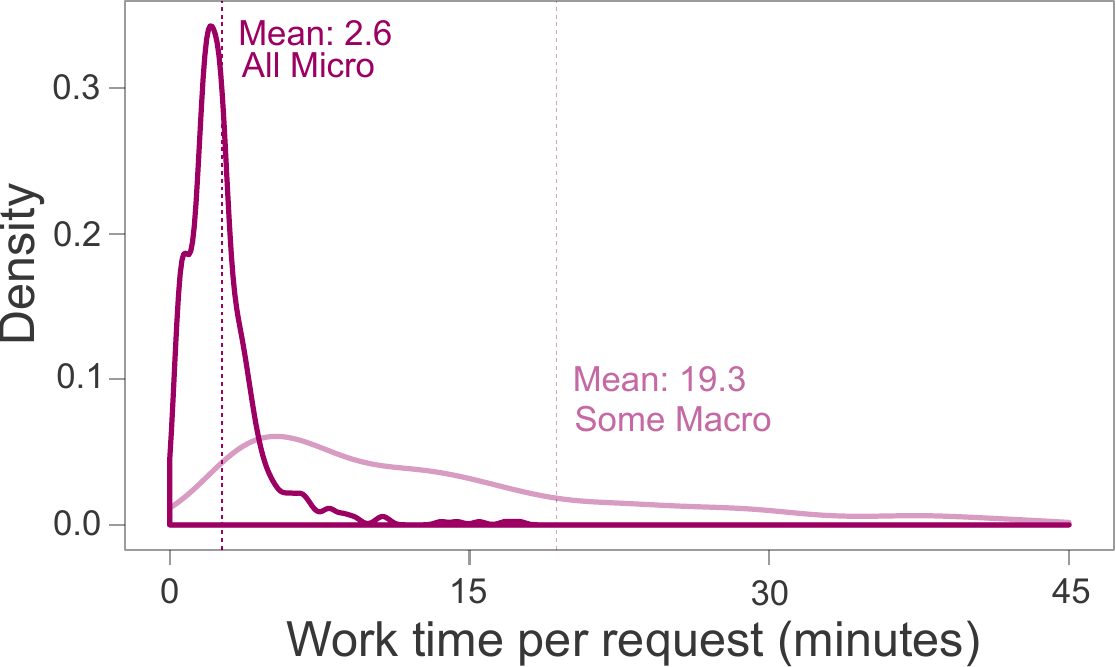}
  \caption{Non-parametric density estimates (Gaussian kernel) of the cumulative work time per request in Study 3b. Requests that did not require macrotasks averaged 2.6 minutes of work time, while requests needing macrotasks averaged 19.3 minutes.}~\label{fig:WorkTimePerRequest}
\end{figure}

Over the course of the three studies, Calendar.help was used by over one hundred people to schedule thousands of meetings. Table~\ref{tab:studies} shows a summary of high-level usage statistics for each study. Starting in Study 3b, we introduced a thorough instrumentation system to measure and record detailed real-time system events and usage metrics. For this reason, we focus the analysis of usage statistics presented below on Study 3b. During this time, 178 participants delegated 1,894 meetings to their Calendar.help assistant. These involved a total of 1,981 unique meeting invitees, and 15,659 emails sent and received by Cal. Most of these meeting requests (82\%, or 1,626) successfully ended in an meeting being scheduled. The remaining 268 requests that did not end in a scheduled meeting were, for one reason or another, cancelled by the organizer prior to completion. These were typically ``test'' requests or recruiters reaching out to people who might have not been interested in a new job.

User engagement was strong and persistent over the course of the field study (see Figure~\ref{fig:UsersPerWeek}). At peak usage in Study 3b, which was just before a major U.S. holiday, over 250 unique users and invitees interacted with the system per week. Although participants in Studies 1 and 2 were compensated for their participation, participants in Study 3 were not. We interpret participants' continuous, and self-motivated engagement as an indication of the utility of Calendar.help, and as signal of Calendar.help being a part of their work routine. 

Of the 178 participants with at least one scheduling requests, 95 of them (53\%) made at least two requests, 75 (42\%) made at least three requests, 57 (32\%) made at least five requests, and 35 (20\%) made ten or more scheduling requests. 
Table~\ref{tab:TopUsers} shows the ten heaviest users of Calendar.help. The top one was part of the study for 13 weeks, and scheduled 180 meetings, averaging 13 meeting requests per week.

Participants used Calendar.help primarily to schedule two-person meetings (i.e. one-on-one), with this type of meeting accounting for 84\% of all requests in Study 3b. This may have been influenced by the significant number of recruiters in the study, who typically scheduled screening calls with job candidates. However, 15\% of meetings involved three or more attendees, up to a maximum of eleven attendees in one request. During the initial Wizard of Oz study (Study 1), we observed that people often scheduled meetings with no invitees (i.e. appointments or reminders). However, this may be an artifact of either getting to know the system or participants' desire to meet the usage requirements, as these amounted to just 0.74\% of all meetings in Study 3b, when most participants were already familiar with the system and they were not required to use the system a specific number of times.

\subsection{System Efficiency}
In Study 3b, 39\% of the requests Calendar.help received were completed entirely within the microtasking workflows of Tiers 1 and 2, never escalating to a macrotask. The remaining 61\% of requests were partially processed in Tiers 1 and 2, requiring some intervention by a macrotask worker at some point in the request life-cycle. Table~\ref{table:macro-creation-reasons} lists the most common reasons for such escalations, including receiving multiple unexpected email responses from an attendee (32\%), failure for the attendees to agree on a time (27\%), and timing-out while awaiting an attendee's response (26\%). 


As one might expect, macrotasks were more time consuming than microtasks. 
Figure~\ref{fig:WorkTimePerRequest} shows density function estimates for the cumulative amount of work time spent per request, broken down into requests that required macrotasks and those that did not. 
On average, requests that did not require macrotasks took only 2.6 minutes of cumulative work time. Those requiring macrotasks took 19.3 minutes, with 15.93 minutes spent on average on macrotasks and 3.4 minutes on microtasks. Note that these 3.4 minutes do not represent wasted effort. 
After escalation to Tier 3, the information extracted from Tiers 1 and 2 is passed on to the macrotask worker by prepopulating certain fields in their user interface, which saves them time.
Requests requiring macrotasks are likely to be inherently more complex and difficult to schedule. This might explain why even the cumulative time in microtasks for requests requiring Tier 3 is higher than the time spent in microtasking for requests that did not require Tier 3 (3.4 versus 2.6 minutes). 
Nonetheless, this supports the hypothesis that task execution via microtasking workflows is more efficient than execution with macrotasking. 

\begin{table}
\centering \small
\begin{tabular}{p{2.6in} r}
\toprule
\bfseries Reason & \bfseries Frequency\\ \midrule
Multiple or out of bound email responses from an attendee.	 &	32\% \\ 
None of the proposed times were acceptable to everyone.  & 27\% \\ 
Timed-out while waiting for a response from an attendee.	& 26\% \\ 
Other / unknown / not instrumented. & 14\% \\ 
Manual (worker) escalation in processing ballot response. & 8\% \\ 
Cannot access organizer's calendar. & 7\% \\ 
Manual (worker) escalation in proposing meeting times. & 7\% \\ 
Manual (worker) escalation in determining attendees.	& 2\% \\
\bottomrule
\end{tabular}
\caption{Reasons for escalation to macrotasks in Study 3b. Note, these are not mutually exclusive: multiple reasons can trigger macrotasks in parallel.}
\label{table:macro-creation-reasons}
\end{table}

\subsection{Boosting Efficiency: Moving from Tier 3 to Tier 2}
As we noted, 
the system architecture leverages historical data to increase task processing efficiency over time. Here we give examples of how, using data collected over the course of the three studies, we identified new ways to boost efficiency by designing additional microtask workflows to prevent requests from escalating to macrotasks.

One way of conceptualizing the system architecture in the Wizard of Oz of Study 1 is as being entirely macrotasks-driven. For each request in Study 1, when the assistant received a new email, the worker had to analyze the request data associated with that email in its entirety to decide what actions to take, without any structured support from the system to guide them. 
In contrast, by Study 3b, 39\% of requests were handled entirely by structured microtasking workflows. Moving from 0\% to 39\% required continual observations of interaction patterns between the assistant and invitees, and incremental improvements to the design to encode the growing scheduling domain knowledge into workflows.

In early studies, for example, we noticed that in phone meetings, some people wanted the invitees' phone numbers on the meeting invitations so they could initiate the call. Prior to having a structured workflow to handle this scenario, these requests would go to a macrotask, where the workers could email the invitees and ask them for their phone numbers.
After observing this scenario frequently enough, we designed a workflow that could capture this interaction. Several efficiency-improvement observations like this one, led to the workflows of Study 3b.

Despite this progress, it's clear from the Study 3b data that additional improvements could be made.
Table~\ref{table:macro-creation-reasons} shows common reasons for macrotask escalation.
As we continue on this trajectory of increasing system efficiency, we look for structural commonalities in escalations that we can exploit for workflow design improvements. 


\subsection{Boosting Efficiency: Moving from Tier 2 to Tier 1}
Data from Study 3b also revealed new ways to improve the automatic execution of microtasks. 
Efforts to introduce automation into the system fall into two categories: (1) automatically handling repeatable, routine tasks in the workflow; and (2) replacing or simplifying the existing microtasks requiring worker's time and attention. 
We present three specific automation improvements in these areas, which are based on a systematic analysis of the data obtained from Study 3b. 
These improvements were not deployed in the field at the time of writing this paper. 

\begin{table}[t]
\centering \small
\begin{tabular}{l l l}
\toprule
\bfseries Feature Type & \bfseries Words & \bfseries Examples\\
\midrule
Days & {Monday}, {Tuesday}, etc. & {Let's do it on Monday.}\\
Ordinal words & {first}, {second}, etc. & {Only the first option.}\\
Logical words & {and}, {or}, etc. & {Monday and Tuesday.}\\
Quantifiers & {all}, {any}, {none}, etc. & {All options work.}\\
\bottomrule
\end{tabular}
\caption{Feature categories and examples for predicting ballot responses.}
\label{table:ballot-response-feature-sets}
\end{table}

\subsubsection{Automated Reminders and Cancellations}
If an invitee fails to respond to a ballot email and a second automated reminder email sent some time later, the workflow engine will, after a reasonable amount of time has passed, escalate the request to an expert worker in macrotasks to try to resolve the unresponsiveness. 
Based on the analysis of the reasons for macrotask escalation shown in Table~\ref{table:macro-creation-reasons}, we observed such scenarios to be the third most common reason for escalation, accounting for 26\% of macrotask escalations. 

Since the automated reminder emails successfully yield an estimated 15\%-22\% additional ballot responses, we decided to implement a second automated reminder, for invitees who still are unresponsive after receiving the first one.  
We expect this reminder to yield an additional 5\%-7\% responses without having to involve macrotask workers to manually write such reminders. 


Despite the automated and manual reminders, and other attempts by the macrotask workers to elicit responses, we discovered that 14\% of all the requests never receive a response and are eventually ignored. 
Based on these results, we introduced an automated cancellation of the request in situations where no response has been received after the second automated reminder. 
Before a cancellation happens, however, a warning is automatically sent to the meeting organizer, giving them a chance to intervene.
Unless the meeting organizer decides to keep the meeting, it is automatically cancelled without requiring additional effort from any worker.

\subsubsection{Extraction of Ballot Selection}
The ballot selection microtask maps the given set of ballot time options and the invitee's ballot email response to the \emph{subset} of choices the invitee selects from the time options. 
The options have a precise structural form, with attributes like the day, time, date, and time zone. 
Users either mark their choices directly by referring to any option using its attributes (e.g. ``Monday at 2pm works for me'') or indirectly using different language constructs (e.g. ``any time after 4pm.'') 

To automate this task, one could use semantic-parsing to map utterances onto logical constructs that indicate which of the options have been selected \cite{artzi2013semantic}. Instead, we explore a much simpler approach by recasting the problem to multiple binary classification tasks on each ballot option, assuming the invitee independently selects one or many choices from the presented options. 

Formally, we can define the task as follows: Every response to a ballot $i$ with $K$ options is a set of 3-tuples for $k=1 \dots K$, $\{k : (r_i, o_i^k, s_i^k)\}$, where $r_i$ is the text response from the user, $o_i^k$ is the $k^{th}$ option in the $i^{th}$ ballot and $s_i^k$ is a binary variable indicating whether this option is selected by the user or not, as given by workers who executed the microtask. We create a binary indicator feature vector for every pair $r_i$ and $o_i^k$, by checking for the presence of hand tuned dictionary words in the response. The feature vector serves as an input to a binary classifier and $s_i^k$ serves as the supervision. Table~\ref{table:ballot-response-feature-sets} shows a sample of different words in broad feature categories in the hand-tuned dictionary.

We selected a sample of ballot responses from microtask output data that had been manually executed in Study 3a and 3b, using 80\% to train the classifier and 20\% to test it. Every response was converted to lowercase and tokenized into words to extract the binary features. We evaluated the model using two metrics: accuracy in predicting each individual ballot choice independently and accuracy in predicting the exact subset of choices on the ballot. We compared Calendar.help's trained classifier with a trivial baseline classifier which predicts the most frequent label in our training data. Table \ref{table:ballot-response-accuracy} shows the results of this comparison. 

This illustrates that data from manually executed microtasks can be used to train machine learning models that can be used to automate the microtask. Even though this model was developed offline after the study was concluded, it's possible to imagine a more sophisticated model operating on live data, which completely automates the ballot response task in high confidence scenarios and falls back to manual execution when its confidence is low.

\begin{table}[t]
\centering \small
\begin{tabular}{l r r}
\toprule
  & \bfseries On individual choices & \bfseries On ballots\\ 
\midrule
 \bfseries Baseline & 63.78\% & 10.94\% \\ 
 \bfseries Logical Classifier & 87.81\% & 73.2\% \\
\bottomrule
\end{tabular}
\caption{Accuracy of the trained logistic regression model for predicting selected ballot choices. The baseline predicts the most common response.}
\label{table:ballot-response-accuracy}
\end{table}

\subsubsection{Extraction of Times from Request}
To explore how extracting time expressions from emails can be used to automated microtasks, we used the SUTime\cite{chang2012sutime} tool, part of the Stanford CoreNLP library,\footnote{\url{http://stanfordnlp.github.io/CoreNLP/}} to extract and interpret time expressions within the user's request email. Because Calendar.help receives a request at any stage of a person's ongoing email conversation with other people, there may be multiple valid time expressions within the thread, not all of which are relevant for scheduling purposes. 
This makes full automation of this task hard. Instead of building a learning system as a first step, our approach was to use heuristic rules to obtain the duration and time of the meeting. 

We used the following heuristics: (1) Use only the duration and date categories from SUTime\footnote{SUTime marks time expressions by categorizing them into 4 categories: time, date, category and set.}, (2) For long email threads, select only time expressions from the latest email, and (3) If multiple time expressions are found, choose the expression closest to the mention of the assistant's name, ``Cal,'' or the first one if Cal is not mentioned.

By employing these heuristics on a sample of 994 requests from Study 3a and 3b, we achieved 62\% accuracy in predicting the duration of the meeting and 39\% accuracy in predicting the meeting time. 

Even though it is encouraging to see the usefulness of these heuristics, one can argue that a system that learns meeting times and durations based on contextual features from emails would outperform a heuristics-based one. However, such a learning system would first require many annotated examples marking the correct text corresponding to meeting time and duration. 
To gather more data, and as an intermediate step for using fully automated predictions in production, we used all the extracted times from a request as suggestions for a microtask worker, with the intention that these could assist the worker. 
The microtask worker can mark if one or all of the suggested times are correct. Designing \emph{assisted microtasks} in this way helps to improve worker efficiency in the short term, and provides the required annotation that a learning system can use in the future.


\subsection{The User Experience}
Over the course of the three studies, we heard from users through interviews and email conversations. Here we aggregate common themes and anecdotes of the experience of having a virtual scheduling assistant.
All participants' names have been changed to respect their privacy. 

\subsubsection{Saving Time through Delegation}
Users reported feeling that they were saving time and being more productive because they were delegating their scheduling tasks to Cal. One direct way this delegation saved people time was by reducing the number of emails they had to read and write. As one participant said: 
\begin{quote}
``Everyone writes so many emails every day. There's several emails confirming a meeting. It's nice to just not to have to do that.'' 
\end{quote}

Another user was more precise, estimating that the burden of scheduling a single meeting was about three to five minutes, time that could be saved by delegating the task to Cal:
\begin{quote}
``Each time with Cal, I'm saving 3--5 minutes. In the process of that, it could save me 2--3 hours of work [per week], probably.''
\end{quote}



Calendar.help sent and received over twenty thousand emails on behalf of its users across the three studies. 

In addition to the time saved in reading and writing email, people specifically liked the fact that Calendar.help tracked the invitees' responses and followed up with them if they did not reply. Some users indicated that this was the biggest burden for them, as many people tend to be unresponsive to invitations.
\begin{quote}
``I have at least three to five [meetings] in a week and it takes a long time to schedule with anyone because everyone is either unresponsive or they'll just give me a whole bunch of different times.''
\end{quote}

However, there was a delicate balancing act between being helpful and being annoying in optimizing the parameters of the automated follow-up logic. Sometimes the system fell on the wrong side of this line. In one case, an invitee (who was a potential job candidate) complained to the meeting organizer (a recruiter) that Cal was being a little too pushy. The organizer had to do some (light-hearted) damage control:
\begin{quote}
``That Cal sure is pushy! Don't worry, I made sure to share this feedback with his manager [meaning our team]. The last thing I want to do is have our scheduling assistant annoy a person we are trying to recruit.''
\end{quote}

We mitigated some of these concerns by ensuring follow-up emails were only sent out during business hours, but additional optimization might be needed.

\subsubsection{Introducing the Assistant}

Considerable variation was observed in how people introduced Cal to their meeting invitees. In most cases, Cal was mentioned explicitly as an `assistant' or a `scheduling assistant' that would take care of setting up the meeting:
\begin{quote}
``I am also including my scheduling assistant, Cal, in this thread. Cal will schedule a date/time for us to chat further.'' 
\end{quote}

Often, these requests would include additional instructions or prompts for Cal to follow:
\begin{quote}
``Adding Cal for scheduling -- Cal, can you find Ariana and I an hour to talk sometime soon?''
\end{quote} 

In other cases, the organizers didn't explicitly introduce Cal as an assistant, and instead addressed him directly, making his subordinate role as an assistant implicit in their request:
\begin{quote}
``Cal, can you please help calendar some time for us? Please make it a Skype meeting.''
\end{quote}

Cal is also often added onto existing email threads when scheduling needs arise, even as other human participants of the thread are dropped off:
\begin{quote}
``I think we can drop Greg to BCC; adding Cal from my side. Cal, can you work with Kaitlin to find 60 minutes for Todd and I sometime next week, preferably later in the week when I'm back on EST?''
\end{quote}

In the last case, one can argue that Cal was sometimes treated as an equal participant of the conversation, indistinct from the human participants. 


In other cases, organizers clearly introduced Cal as distinctly non-human. For example here, Cal is described as a `virtual assistant':
\begin{quote}
``I'm cc'ing my virtual assistant (Cal) to help with scheduling. Please let us know when might be a good time for a call.''
\end{quote}

Similarly, other users used the term `scheduling bot', `digital assistant', or simply `bot.' Additional work is needed to explore what impact, if any, these different ways of introducing Cal has on the resulting scheduling interactions between Cal and the attendees. 

\subsubsection{Social Status}

In many cases, invitees thought of Cal as a real person.
\begin{quote}
``A lot of people thought I had an actual, like, live assistant, which is actually pretty funny. But I think you guys are on the right track by giving it a human element, because it definitely comes across. It does not seem very automated at all. So that's a nice, personal touch to it.''
\end{quote}

This `human element' of Cal had interesting social implications for the organizer. Because personal assistants are typically reserved for senior roles, such as executives, some participants perceived a status increase while using Cal.
\begin{quote}
``The other best thing [about Cal] was people thought it was really cool and they're like, `What is this Margret? How come you have an electronic assistant? How can I get one?' and I'm thinking, `How can I keep him?' Right? Because it was really cool.''
\end{quote}

This status bump was not always welcome. One participant 
was in an outreach role within his company that required him to cultivate a network of peers in other companies, whom he met with regularly. He felt that having an assistant might add friction to these professional relationships, as he didn't wish to be elevated above the people he meets with. He sometimes referred to Cal as his `colleague' instead of his assistant, to mitigate this concern. 

People were also conscious of the relative status of the person they were meeting with when using Calendar.help, typically avoiding Calendar.help for setting up meetings with people who are more senior to them in their organization.





\subsubsection{Pushing Boundaries}

Because interaction with Calendar.help was natural and open-ended, users were not always certain about what the system could do for them. Sometimes the system's capabilities surprised them by performing something they did not anticipate. 
For example, one user who thought Cal could only handle the initial scheduling was surprised when rescheduling request was addressed:
\begin{quote}
``Loving Cal so far! It even took care of a reschedule nicely and scheduled it for 2pm the same day, it was magical.''
\end{quote}


Other times the requests users made stretched beyond what Cal could handle.
For example, another user asked Cal to do some basic research for him in preparation for his meeting:
\begin{quote}
``I asked him a question. I was actually looking for some very important information about the federal law and he came back and said he [couldn't] do that quite yet but if he could, he would.''
\end{quote}

Although research tasks are beyond even the broad scope of Calendar.help, other requests were at the boundary of what Calendar.help was designed for. For example, some users wanted to have a summary of the meetings scheduled for the week, while others wanted help getting a PIN for a conference call from their IT department.

As these requests indicate, even the domain of meeting scheduling can involve many different processes in different work environments. 
Requests like these represent potential areas for growth, but also indicate the need for explicit guidelines for the 
macrotask workers to scope their responsibilities,
especially because users are likely to test the boundaries.

\section{DISCUSSION}



By looking at how hundreds of Calendar.help subscribers used the system to schedule thousands of meetings, we showed that it is possible to handle a broad range of calendaring needs efficiently and robustly using a novel three-tiered architecture. By seamlessly combining automation, microtasking, and macrotask execution,
we were able to take the approach used in Wizard of Oz studies \cite{kelley1984iterative} out of the lab and into the wild.
Study 1 was entirely Wizard of Oz based, but as we learned common tasks through continued usage we were able to decompose these aspects into microtasks for manual execution, and eventually -- as we collected enough usage data -- for automation.
We believe that this progressive \emph{Wizard in the Wild} design approach can be applied to other complex tasks in a way that will enable robust intelligent systems to support people before they are more thoroughly automated.



Calendar.help's three-tiered architecture could be seen as an intermediate stage on the long and technically challenging path to full artificial intelligence, at least for narrow domains like scheduling.
Despite advances in natural language processing, extracting information from free-text is still error-prone and algorithms often make mistakes that seem trivial to humans. Improving automation requires large amounts of high-quality training data. The Wizard-in-the-Wild approach offers a platform to collect and annotate this data in realistic settings, while simultaneously providing a service to end users. 

To support the transition to automation Calendar.help employs \emph{assisted microtasks}, which use automation to suggest answers to microtasks but require worker verification.
Assisted microtasks help workers perform tasks quicker, as they typically require only yes/no verifications. They also help minimize mistakes that would occur during automation, and enable the collection of valuable training data that allows for a gradual ramping-up of automation over time.

In general, automation helps people avoid having to complete routine and tedious work, which allows them to devote time to more complex, creative work. Our goal with Calendar.help is not to replace personal assistants, but rather to support a democratization of the service, allowing more people -- including personal assistants -- to worry less about scheduling. Further, Calendar.help enables workers who chose to remain focused on scheduling to become more efficient by working through the system to provide scheduling expertise to a larger set of people than currently possible. 

It  remains an open question as to whether complete automation is a desirable outcome.
Even for scheduling tasks that are easily automated we sometimes observed that the Calendar.help macrotask approach provided users with unexpected value.
Cal informed one user, for example, of a typo in her email signature during a routine scheduling interaction.
The benefits of rich human interaction can be lost if rapid movement towards cost-effective automation is prioritized too early. 
We may find that some users in the automated future wish to keep these human-in-the-loop elements, perhaps at a premium.

\subsection{Architecture Limitations}
Despite the benefits provided by Calendar.help's three-tiered architecture, there are some limitations. For one, workflow design is not trivial. Significant resources are needed to fine-tune the structured microtasks so that they are resilient to worker errors and can accommodate a diverse pool of workers.
Further, debugging workflows can be complex, as they can have numerous execution paths that must be considered. Because of this, workflows may be better-suited for problems with a limited number of states. Scheduling, though complex, is a narrowly scoped problem, making it suitable for this kind of architecture. Much more open ended scenarios, such as with a general-purpose conversational assistant, may be difficult to support and implement with a workflow-based approach.

Additionally, the cost of having humans in the loop from the beginning of the design process is potentially prohibitive for some system designers. This is particularly salient if the goal is to create real-time systems where low latency is essential.
Since the architecture depends on a layer of macrotask workers, these workers require training and expertise to execute the macrotasks. Training requires time and resources that need to be taken into account during the design of the system. Additionally, this training could impact the ability to quickly scale the system up while maintaining quality.

Calendar.help's architecture also means that the end user never knows exactly who -- or what -- they are communicating with when they email Cal.
It is important that the system transparently communicate that multiple humans may be involved in processing requests.
Although participants never expressed a concern about privacy, hybrid intelligence systems like Calendar.help must ensure they do not inappropriately expose personal information. 
Calendar.help addressed this by having workers sign non-disclosure agreements and by designing microtasks to include only the information needed to complete the task.
There may also be ways to intentionally break tasks down such that potentially identifiable information is separated into different microtasks and is therefore unidentifiable (or more difficult to identify) in isolation \cite{laseckipreserving}.
The structured workflows and automated support for microtasks used by Calendar.help could also be harnessed by the users themselves to more easily complete their own scheduling tasks through selfsourcing \cite{teevan2014selfsourcing}.

\subsection{System Limitations}
We observed some usability limitations with Calendar.help. For complex meetings, particularly those with a large number of invitees, it was sometimes difficult for the assistant to converge on a meeting time without sharing calendar information among all invitees. In these cases shared polling approaches may be more efficient, since people will sometimes change their constraints based on awareness of others \cite{Reinecke2013doodle}.

Although meeting organizers saved considerable time by delegating their scheduling to Calendar.help, meeting invitees still had to handle the burden of manually responding to Cal with their availability. In some cases, this might be frustrating for them, as they have to cross reference the possible time options with their online calendars, which requires coordination. Ideally, Calendar.help should also be a smooth experience for invitees. We also saw friction in using Calendar.help for setting up meetings for people within teams in a large enterprise with complex demands for meeting room reservations, where existing solutions for sharing internal calendar information for booking meetings were already in place.

\section{CONCLUSION}
In this paper, we introduce Calendar.help, a system that provides fast, efficient, personalized scheduling through structured workflows. Users communicate with Cal as if it were a personal assistant, but only the most challenging calendaring tasks emailed to the system are done by trained assistants. Instead, common calendaring tasks are broken down and completed by microtask workers or automated processes. We described an iterative approach used to develop Calendar.help, and shared what we learned about the transition between the macro-calendaring task and structured calendaring microtasks by deploying the system to hundreds of users.
Hybrid intelligence systems like Calendar.help, successfully transition between macrotasks and microtasks and can enable people to seamlessly delegate a wide range of complex tasks to agents powered by machines and people of varying expertise.

\bibliographystyle{SIGCHI-Reference-Format}
\bibliography{projectch}


\begin{thebibliography}{00}


\ifx \showCODEN    \undefined \def \showCODEN     #1{\unskip}     \fi
\ifx \showDOI      \undefined \def \showDOI       #1{{\tt DOI:}\penalty0{#1}\ }
  \fi
\ifx \showISBNx    \undefined \def \showISBNx     #1{\unskip}     \fi
\ifx \showISBNxiii \undefined \def \showISBNxiii  #1{\unskip}     \fi
\ifx \showISSN     \undefined \def \showISSN      #1{\unskip}     \fi
\ifx \showLCCN     \undefined \def \showLCCN      #1{\unskip}     \fi
\ifx \shownote     \undefined \def \shownote      #1{#1}          \fi
\ifx \showarticletitle \undefined \def \showarticletitle #1{#1}   \fi
\ifx \showURL      \undefined \def \showURL       #1{#1}          \fi

\bibitem{angeli2012parsing}
{Gabor Angeli}, {Christopher~D Manning}, {and} {Daniel Jurafsky}. 2012.
\newblock \showarticletitle{Parsing time: Learning to interpret time
  expressions}. In {\em Proceedings of the 2012 Conference of the North
  American Chapter of the Association for Computational Linguistics: Human
  Language Technologies}. Association for Computational Linguistics, 446--455.
\newblock


\bibitem{artzi2013semantic}
{Yoav Artzi}, {Nicholas FitzGerald}, {and} {Luke~S Zettlemoyer}. 2013.
\newblock \showarticletitle{Semantic Parsing with Combinatory Categorial
  Grammars.}
\newblock {\em ACL (Tutorial Abstracts)\/}  {3} (2013).
\newblock


\bibitem{bardram2005web}
{Jakob~E. Bardram} {and} {Claus Bossen}. 2005.
\newblock \showarticletitle{A web of coordinative artifacts: collaborative work
  at a hospital ward}. In {\em Proceedings of the 2005 International ACM
  SIGGROUP Conference on Supporting Group Work}. ACM, 168--176.
\newblock
\showURL{%
\url{http://dx.doi.org/10.1145/1099203.1099235}}


\bibitem{birkinshaw2013make}
{Julian Birkinshaw} {and} {Jordan Cohen}. 2013.
\newblock \showarticletitle{Make time for the work that matters}.
\newblock {\em Harvard Business Review\/} (01 September 2013).
\newblock
\newblock
\shownote{\url{https://hbr.org/2013/09/make-time-for-the-work-that-matters}.}


\bibitem{brzozowski2006grouptime}
{Mike Brzozowski}, {Kendra Carattini}, {Scott~R. Klemmer}, {Patrick Mihelich},
  {Jiang Hu}, {and} {Andrew~Y. Ng}. 2006.
\newblock \showarticletitle{groupTime: Preference based group scheduling}. In
  {\em Proceedings of the SIGCHI Conference on Human Factors in Computing
  Systems}. ACM, 1047--1056.
\newblock
\showURL{%
\url{http://dx.doi.org/10.1145/1124772.1124929}}


\bibitem{chang2012sutime}
{Angel~X. Chang} {and} {Christopher~D. Manning}. 2012.
\newblock \showarticletitle{SUTime: A library for recognizing and normalizing
  time expressions}. In {\em LREC}. 3735--3740.
\newblock


\bibitem{chilton2013cascade}
{Lydia~B. Chilton}, {Greg Little}, {Darren Edge}, {Daniel~S. Weld}, {and}
  {James~A. Landay}. 2013.
\newblock \showarticletitle{Cascade: Crowdsourcing taxonomy creation}. In {\em
  Proceedings of the SIGCHI Conference on Human Factors in Computing Systems}.
  ACM, 1999--2008.
\newblock
\showURL{%
\url{http://dx.doi.org/10.1145/2470654.2466265}}


\bibitem{ducheneaut2001mail}
{Nicolas Ducheneaut} {and} {Victoria Bellotti}. 2001.
\newblock \showarticletitle{E-mail as habitat: an exploration of embedded
  personal information management}.
\newblock {\em Interactions\/} {8}, 5 (2001), 30--38.
\newblock
\showURL{%
\url{http://dl.acm.org/citation.cfm?doid=382899.383305}}


\bibitem{Ehrlich1987social}
{Susan~F. Ehrlich}. 1987a.
\newblock \showarticletitle{Social and psychological factors influencing the
  design of office communication systems}. In {\em Proceedings of the SIGCHI/GI
  Conference on Human Factors in Computing Systems and Graphics Interface} {\em
  (CHI '87)}. 323--329.
\newblock
\showURL{%
\url{http://dx.doi.org/10.1145/29933.275651}}


\bibitem{ehrlich1987strategies}
{Susan~F. Ehrlich}. 1987b.
\newblock \showarticletitle{Strategies for encouraging successful adoption of
  office communication systems}.
\newblock {\em ACM Transactions on Information Systems (TOIS)\/} {5}, 4 (1987),
  340--357.
\newblock
\showURL{%
\url{http://dx.doi.org/10.1145/42196.42198}}


\bibitem{etzioni2005unsupervised}
{Oren Etzioni}, {Michael Cafarella}, {Doug Downey}, {Ana-Maria Popescu}, {Tal
  Shaked}, {Stephen Soderland}, {Daniel~S Weld}, {and} {Alexander Yates}. 2005.
\newblock \showarticletitle{Unsupervised named-entity extraction from the web:
  An experimental study}.
\newblock {\em Artificial intelligence\/} {165}, 1 (2005), 91--134.
\newblock


\bibitem{grudin1988cscw}
{Jonathan Grudin}. 1988.
\newblock \showarticletitle{Why CSCW applications fail: Problems in the design
  and evaluation of organizational interfaces}. In {\em Proceedings of the 1988
  ACM Conference on Computer-Supported Cooperative Work} {\em (CSCW '88)}.
  85--93.
\newblock
\showURL{%
\url{http://dx.doi.org/10.1145/62266.62273}}


\bibitem{wired_facebookM}
{Jessi Hempel}. 2015.
\newblock \showarticletitle{Facebook launches M, its bold answer to Siri and
  Cortana}.
\newblock {\em Wired\/} (2015).
\newblock
\newblock
\shownote{\url{https://www.wired.com/2015/08/facebook-launches-m-new-kind-virtual-assistant/}.}


\bibitem{horvitz1999principles}
{Eric Horvitz}. 1999.
\newblock \showarticletitle{Principles of mixed-initiative user interfaces}. In
  {\em Proceedings of the SIGCHI Conference on Human Factors in Computing
  Systems}. ACM, 159--166.
\newblock
\showURL{%
\url{http://dx.doi.org/10.1145/302979.303030}}


\bibitem{bloomberg_chatbots}
{Ellen Huet}. 2016.
\newblock \showarticletitle{The humans hiding behind the chatbots}.
\newblock {\em Bloomberg Technology\/} (2016).
\newblock
\newblock
\shownote{\url{http://www.bloomberg.com/news/articles/2016-04-18/the-humans-hiding-behind-the-chatbots}.}


\bibitem{Cheng2015break}
{Shamsi T.~Iqbal Justin~Cheng, Jaime~Teevan} {and} {Michael~S. Bernstein}.
  2015.
\newblock \showarticletitle{Break it down: A comparison of macro- and
  microtasks.}. In {\em Proceedings of the 33rd Annual ACM Conference on Human
  Factors in Computing Systems} {\em (CHI '15)}. 4061--4064.
\newblock
\showURL{%
\url{http://dx.doi.org/10.1145/2702123.2702146}}


\bibitem{hybridintelligence}
{Ece Kamar}. 2016.
\newblock \showarticletitle{Hybrid Intelligence and the Future of Work}. In
  {\em Productivity Decomposed: Getting Big Things Done with Little Microtasks
  Workshop} {\em (CHI 2016)}.
\newblock
\showURL{%
\url{http://research.microsoft.com/en-us/um/people/eckamar/papers/HybridIntelligence.pdf}}


\bibitem{kamar2012combining}
{Ece Kamar}, {Severin Hacker}, {and} {Eric Horvitz}. 2012.
\newblock \showarticletitle{Combining human and machine intelligence in
  large-scale crowdsourcing}. In {\em Proceedings of the 11th International
  Conference on Autonomous Agents and Multiagent Systems-Volume 1}.
  International Foundation for Autonomous Agents and Multiagent Systems,
  467--474.
\newblock


\bibitem{Reinecke2013doodle}
{Abraham Bernstein Michael~Naf Katharina~Reinecke, Minh Khoa~Nguyen} {and}
  {Krzysztof~Z. Gajos}. 2013.
\newblock \showarticletitle{Doodle around the world: Online scheduling behavior
  reflects cultural differences in time perception and group decision-making}.
  In {\em Proceedings of the 2013 Conference on Computer Supported Cooperative
  Work}. 45--54.
\newblock
\showURL{%
\url{http://dx.doi.org/10.1145/2441776.2441784}}


\bibitem{Kellermann2009privacy}
{Benjamin Kellermann} {and} {Rainer Bohme}. 2009.
\newblock {\em Privacy-Enhanced Event Scheduling}.
\newblock IEEE Computer Society. 52--59 pages.
\newblock


\bibitem{kelley1984iterative}
{John~F. Kelley}. 1984.
\newblock \showarticletitle{An iterative design methodology for user-friendly
  natural language office information applications}.
\newblock {\em ACM Transactions on Information Systems (TOIS)\/} {2}, 1 (1984),
  26--41.
\newblock
\showURL{%
\url{http://dx.doi.org/10.1145/357417.357420}}


\bibitem{kittur2013future}
{Aniket Kittur}, {Jeffrey~V. Nickerson}, {Michael Bernstein}, {Elizabeth
  Gerber}, {Aaron Shaw}, {John Zimmerman}, {Matt Lease}, {and} {John Horton}.
  2013.
\newblock \showarticletitle{The Future of Crowd Work}. In {\em Proceedings of
  the 2013 Conference on Computer Supported Cooperative Work} {\em (CSCW '13)}.
  ACM, New York, NY, USA, 1301--1318.
\newblock
\showISBNx{978-1-4503-1331-5}
\showDOI{%
\url{http://dx.doi.org/10.1145/2441776.2441923}}


\bibitem{laseckipreserving}
{Walter~S. Lasecki}, {Mitchell Gordon}, {Jaime Teevan}, {Ece Kamar}, {and}
  {Jeffrey~P Bigham}. 2015.
\newblock \showarticletitle{Preserving Privacy in Crowd-Powered Systems}.
\newblock  (2015).
\newblock


\bibitem{lasecki2013chorus}
{Walter~S. Lasecki}, {Rachel Wesley}, {Jeffrey Nichols}, {Anand Kulkarni},
  {James~F Allen}, {and} {Jeffrey~P. Bigham}. 2013.
\newblock \showarticletitle{Chorus: a crowd-powered conversational assistant}.
  In {\em Proceedings of the 26th Annual ACM Symposium on User Interface
  Software and Technology}. ACM, 151--162.
\newblock
\showURL{%
\url{http://dx.doi.org/10.1145/2501988.2502057}}


\bibitem{maes1994agents}
{Pattie Maes} {and} {others}. 1994.
\newblock \showarticletitle{Agents that reduce work and information overload}.
\newblock {\it Commun. ACM} {37}, 7 (1994), 30--40.
\newblock
\showURL{%
\url{http://dx.doi.org/10.1145/176789.176792}}


\bibitem{wired_ai_helps_humans}
{Cade Metz}. 2015.
\newblock \showarticletitle{AI helps humans best when humans help the AI}.
\newblock {\em Wired\/} (2015).
\newblock
\newblock
\shownote{\url{https://www.wired.com/2015/09/ai-helps-humans-best-humans-help-ai/}.}


\bibitem{mitchell1994experience}
{Tom~M. Mitchell}, {Rich Caruana}, {Dayne Freitag}, {John McDermott}, {David
  Zabowski}, {and} {others}. 1994.
\newblock \showarticletitle{Experience with a learning personal assistant}.
\newblock {\it Commun. ACM} {37}, 7 (1994), 80--91.
\newblock
\showURL{%
\url{http://dx.doi.org/10.1145/176789.176798}}


\bibitem{myers2007intelligent}
{Karen Myers}, {Pauline Berry}, {Jim Blythe}, {Ken Conley}, {Melinda Gervasio},
  {Deborah~L McGuinness}, {David Morley}, {Avi Pfeffer}, {Martha Pollack},
  {and} {Milind Tambe}. 2007.
\newblock \showarticletitle{An intelligent personal assistant for task and time
  management}.
\newblock {\em AI Magazine\/} {28}, 2 (2007), 47.
\newblock


\bibitem{retelny2014Expert}
{Daniela Retelny}, {S{\'e}bastien Robaszkiewicz}, {Alexandra To}, {Walter~S.
  Lasecki}, {Jay Patel}, {Negar Rahmati}, {Tulsee Doshi}, {Melissa Valentine},
  {and} {Michael~S. Bernstein}. 2014.
\newblock \showarticletitle{Expert Crowdsourcing with Flash Teams}. In {\em
  Proceedings of the 27th Annual ACM Symposium on User Interface Software and
  Technology} {\em (UIST '14)}. ACM, New York, NY, USA, 75--85.
\newblock
\showISBNx{978-1-4503-3069-5}
\showDOI{%
\url{http://dx.doi.org/10.1145/2642918.2647409}}


\bibitem{sen1998formal}
{Sandip Sen} {and} {Edmund~H. Durfee}. 1998.
\newblock \showarticletitle{A formal study of distributed meeting scheduling}.
\newblock {\em Group Decision and Negotiation\/} {7}, 3 (1998), 265--289.
\newblock
\showURL{%
\url{https://doi.org/10.1145/122831.122837}}


\bibitem{tang2011your}
{John~C. Tang}, {Chen Zhao}, {Xiang Cao}, {and} {Kori Inkpen}. 2011.
\newblock \showarticletitle{Your time zone or mine?: a study of globally time
  zone-shifted collaboration}. In {\em Proceedings of the ACM 2011 Conference
  on Computer-Supported Cooperative Work}. ACM, 235--244.
\newblock
\showURL{%
\url{http://dx.doi.org/10.1145/1958824.1958860}}


\bibitem{teevan2016microwork}
{Jaime Teevan}. 2016.
\newblock \showarticletitle{The future of microwork}.
\newblock {\em XRDS: Crossroads\/} {23}, 2 (2016), 26--29.
\newblock
\showURL{%
\url{https://doi.org/10.1145/3019600}}


\bibitem{teevan2016supporting}
{Jaime Teevan}, {Shamsi~T. Iqbal}, {and} {Curtis von Veh}. 2016.
\newblock \showarticletitle{Supporting Collaborative Writing with Microtasks}.
  In {\em Proceedings of the 2016 CHI Conference on Human Factors in Computing
  Systems} {\em (CHI '16)}. ACM, New York, NY, USA, 2657--2668.
\newblock
\showISBNx{978-1-4503-3362-7}
\showDOI{%
\url{http://dx.doi.org/10.1145/2858036.2858108}}


\bibitem{teevan2014selfsourcing}
{Jaime Teevan}, {Daniel~J. Liebling}, {and} {Walter~S. Lasecki}. 2014.
\newblock \showarticletitle{Selfsourcing personal tasks}. In {\em CHI'14
  Extended Abstracts on Human Factors in Computing Systems}. ACM, 2527--2532.
\newblock
\showURL{%
\url{http://dx.doi.org/10.1145/2559206.2581181}}


\bibitem{Erickson2008assistance}
{Wendy A.~Kellogg Thomas~Erickson, Catalina M.~Danis} {and} {Mary~E. Helander}.
  2008.
\newblock \showarticletitle{Assistance: The work practices of human
  administrative assistants and their implications for it and organizations}.
  In {\em Proceedings of the 2008 ACM Conference on Computer Supported
  Cooperative Work} {\em (CSCW '08)}. 608--618.
\newblock
\showURL{%
\url{http://dx.doi.org/10.1145/1460563.1460658}}


\bibitem{Myllarniemi2014meeting}
{Mikko Raatikainen Terho~Norja Varvana~Myllärniemi, Olli~Korjus} {and} {Tomi
  Männistö}. 2014.
\newblock \showarticletitle{Meeting scheduling across heterogeneous calendars
  and organizations utilizing mobile devices and cloud services}. In {\em
  Proceedings of the 13th International Conference on Mobile and Ubiquitous
  Multimedia} {\em (MUM '14)}. 224--227.
\newblock
\showURL{%
\url{http://dx.doi.org/10.1145/2677972.2678002}}


\bibitem{xu2013convolutional}
{Puyang Xu} {and} {Ruhi Sarikaya}. 2013.
\newblock \showarticletitle{Convolutional neural network based triangular crf
  for joint intent detection and slot filling}. In {\em Automatic Speech
  Recognition and Understanding (ASRU), 2013 IEEE Workshop on}. IEEE, 78--83.
\newblock


\bibitem{zhang2012human}
{Haoqi Zhang}, {Edith Law}, {Rob Miller}, {Krzysztof Gajos}, {David Parkes},
  {and} {Eric Horvitz}. 2012.
\newblock \showarticletitle{Human computation tasks with global constraints}.
  In {\em Proceedings of the SIGCHI Conference on Human Factors in Computing
  Systems}. ACM, 217--226.
\newblock
\showURL{%
\url{http://dx.doi.org/10.1145/2207676.2207708}}


\bibitem{zunino2009chronos}
{Alejandro Zunino} {and} {Marcelo Campo}. 2009.
\newblock \showarticletitle{Chronos: A multi-agent system for distributed
  automatic meeting scheduling}.
\newblock {\em Expert Systems with Applications\/} {36}, 3 (2009), 7011--7018.
\newblock


\end{thebibliography}

\balance

\end{document}